\documentstyle[12pt,epsfig]{article}
\def\fig#1{Fig.~{\ref{#1}}}

\newskip\humongous \humongous=0pt plus 1000pt minus 1000pt
\def\caja{\mathsurround=0pt}
\def\eqalign#1{\,\vcenter{\openup1\jot \caja
        \ialign{\strut \hfil$\displaystyle{##}$&$
        \displaystyle{{}##}$\hfil\crcr#1\crcr}}\,}
\newif\ifdtup

\newcounter{eqnumber}
\renewcommand{\theeqnumber}{\arabic{eqnumber}}
\def\equn{
\refstepcounter{eqnumber}
\eqno({\rm \theeqnumber})
}

\def\gluino{{\tilde g}}
\def\qb{{\bar q}}

\begin{document}

\begin{titlepage}

\begin{flushright}
hep-ph/9606466 \hfill UCLA/96/TEP/21\\
June 27, 1996\\
\end{flushright}

\vskip 2.cm

\begin{center}
{\Large\bf Light Gluinos and Jet Production in $\bar p p$ Collisions} 
\vskip 2.cm

{\large Zvi Bern, Aaron K. Grant, and Andrew G. Morgan}

\vskip 0.5cm

{\it  Department of Physics, University of California at Los
Angeles, CA 90095-1547}
\vskip 3cm
\end{center}

\begin{abstract}
  The impact of a light, long lived gluino on jet production cross
  sections in $\bar p p$ collisions is estimated.  The effect is found
  to be relatively modest, particularly when gluinos are incorporated
  into the parton densities of the proton.  Although a light gluino
  does enhance the production of jets at high $E_T$, the effect is
  insufficient, by itself, to explain the high $E_T$ excess observed
  by CDF.
\end{abstract}

\vfill
\end{titlepage}

\section{Introduction}

In addition to those which have already been observed, the minimal
supersymmetric standard model predicts a plethora of new particles.
Among these new states is the superpartner of the gluon, a color octet
Majorana fermion known as the gluino.  It is generally believed that
the superpartners of the known particles should have masses of order
$10^{2-3}$ GeV, but experimental searches have yet to find any direct
evidence for supersymmetry.  Some authors, however, have argued that a
very light gluino with a mass in the several hundred MeV range is
difficult to exclude on the basis of present data
\cite{mayberefs,Farrar}.

A number of experiments have searched for light gluinos. The
persistently negative results obtained have been used to exclude large
ranges of masses and lifetimes for this particle.  These experiments
include various beam dump experiments \cite{bmdmp}, searches for
exotic particles in various mass and lifetime ranges \cite{stable},
searches for gluino-containing hadrons produced in $\Upsilon$ and
$\chi_b$ decay \cite{Upsilon}, and collider searches \cite{UA1}.
Taken at face value, these experiments exclude all but a few narrow
windows for the light gluino.  It has been argued, however, that the
excluded range may have been overestimated \cite{Farrar,Fupsilon}, and
that sizable allowed regions remain.

Very recently, data from the Fermilab E761 experiment \cite{E761} have
been re-analyzed to derive bounds on the production rate of
gluino-containing baryons with lifetimes in the range 50 to 500
picoseconds.  This experiment placed stringent constraints on the
fraction of gluino-containing baryons that are produced when 800 GeV
protons are incident on a copper target.  The fraction is found to be
less than about $10^{-5}$ for relatively light ($\sim \! 1.7$ GeV)
gluino containing baryons.  However, the acceptance of the experiment
is not sufficient to place as stringent bounds on the production
fraction of heavier ($\sim 2.5$ GeV) supersymmetric baryons.

There is some controversy regarding the exact mass and lifetime
regions allowed for light gluinos. A fairly liberal estimate of the
allowed regions may be found in Ref.~\cite{Farrar}.  It is argued
there that gluinos which hadronize into gluino-gluon bound states of
mass less than about 2 GeV are allowed for certain lifetime ranges,
and that a gluino of mass greater than about 4 GeV is allowed with
certain lifetimes in the range greater than about $10^{-10}$ seconds.
Some of the experiments cited above claim that the allowed regions are
smaller.

The searches discussed above run into potential difficulties either
because of non-perturbative hadronic uncertainties, or because of the
possibility of a long gluino lifetime.  High $E_T$ jet physics has the
potential to skirt around some of these difficulties.  This is mainly
because the essential physics is perturbative, and the final result is
rather insensitive to the mass of the light gluino.  In this letter,
we investigate whether jet production data in $\bar p p$ collisions
can be used to differentiate between the Standard Model and the
Standard Model plus a light, long lived, gluino.

In $ep$ collisions, gluinos do not participate in the hard scattering,
and their effect on the quark distributions has been found to be
negligible \cite{RS,RV}. In contrast, jet production in $\bar p p$
collisions is sensitive to the gluino at leading order in perturbation
theory.  One might suspect the effect on jet cross sections to be
rather large since a single Majorana gluino, in many cases, has the
same effect as three light quarks.  Below, we estimate the effect of a
light gluino on $\bar p p$ jet production rates by calculating the
single-jet-inclusive $E_T$ spectrum.  We will show that competing
effects tend to suppress the effect of a light gluino.

Before proceeding, it is worth noting a few of the assumptions
implicit in our work.  First, we assume that gluinos, if they exist,
are sufficiently long lived that they will hadronize and form jets in
the same manner as other strongly interacting particles.  We
furthermore assume that the missing energy resulting from gluino decay
into photinos or other weakly interacting particles is negligible.  We
also assume that other strongly interacting supersymmetric particles,
such as squarks, are sufficiently heavy that their effects can be
neglected; the presence of a few hundred GeV squark would lead to a
peak in the dijet mass distribution not predicted by Standard Model
physics \cite{ClavelliDiJet}.  (As yet, no such peak has been
established \cite{CDF2JET} in the data.)

The layout of this letter is as follows.  In Sec. \ref{JetSpectrum},
we discuss the effects of a light gluino on the single-jet-inclusive
$E_T$ spectrum.  In Sec. \ref{VsExperiment}, we review the relevant
experimental results, and discuss what can be said about a light
gluino on the basis of our calculation.  Sec. \ref{Summary} concludes.

\section{Light gluinos and the jet $E_T$ spectrum}
\label{JetSpectrum}

In order to estimate the effect of a light gluino on the single-jet
inclusive cross section we have modified the jet Monte Carlo program
JETRAD \cite{GieleGlover} to include gluinos both in the evolution of
$\alpha_s$ and in the hard scattering processes.  Parton densities for
the proton that include a gluino have been obtained from the authors
of Ref.~\cite{RS}. All of the calculations presented here have been
performed to leading order in the QCD coupling parameter, $\alpha_s$,
using a two-loop running of the coupling.  In order to match to CDF
parameters, the computation was performed for transverse energies,
$E_T$, in the range $50~{\rm GeV} < E_T < 450~{\rm GeV}$ and for
pseudorapidities, $\eta$, in the range $0.1<|\eta|<0.7$.  Following
standard choices, the renormalization and factorization scales have
been set equal to one half of the $E_T$ of the leading jet.

Since we are primarily interested in the effect of a gluino on the
shape of the $E_T$ spectrum, leading order perturbation theory is
sufficient for our purposes.  For standard QCD, the main effect of
next-to-leading order corrections is to rescale this cross section by
a nearly constant `$K$-factor' over the entire range of $E_T$
\cite{GMRS}. (Using JETRAD we have verfied this with the above choice of
parameters.)

The light gluino has three main effects on the jet $E_T$ spectrum: it
modifies the running of $\alpha_s$, introduces new states into the
scattering process, and modifies the
Dokshitzer-Gribov-Lipatov-Altarelli-Parisi (DGLAP) \cite{DGLAP}
evolution of the parton densities.  In order to obtain a proper
estimate of the effect we must include all three of these effects as
they are equally important. Below we consider how each of these three
effects alter the theoretical prediction of the single-jet-inclusive
cross section.  We shall find that although the first two tend to
increase the cross section preferentially in the high $E_T$ region,
the inclusion of a gluino in the parton distributions tends to cancel
part of this increase.

The evolution of the coupling constant is controlled by the
beta-function
$$
\beta_g  = {g^3 \over 16 \pi^2} \biggl[ -{11 N_c \over 3} + {2\over 3} n_f
+ {2N_c \over 3} n_\gluino \biggr] + {\cal O}(g^5) \, , 
\equn
$$
where $n_{\! f}$ is the number of quarks, $n_\gluino$ is the number of
gluinos which we take to be either zero or one, and $N_c=3$ is the
number of colors. In running the coupling we follow the standard
prescription that below a particle threshold the effect of that
particle is ignored, but at threshold the contribution to the
beta-function is turned on.  (To maintain consistency with the
conventions of Ref.~\cite{RS} we take the threshold to be twice the
mass of the parton --- our final results are insensitive to this
choice.)  The addition of a light fermionic degree of freedom slows
the running of $\alpha_s$, since it makes the beta function less
negative. Indeed, it has been argued that a light gluino could resolve
the long-standing but small disagreement between measurements of
$\alpha_s$ at low scales ($\sim 5$ GeV) and at high scales ($\sim 91$
GeV) \cite{BB}.

Starting from the measured value of the strong coupling constant in
deep inelastic scattering experiments, a light gluino shifts the value
at the $Z$-resonance from $\alpha_s(M_Z) = 0.110$ to $\alpha_s(M_Z) =
0.122\,$. These values of $\alpha_s$ are dictated by our choice of
parton distribution functions.  Since leading order contributions to
the single-jet-inclusive production rate are proportional to two
powers of $\alpha_s$, this causes about a 17\% rise in the
high $E_T$ end of the spectrum.  (The coupling constant at about 10
GeV is the same with or without gluinos since it is constrained by the
deep inelastic scattering data \cite{DIS}.)

\begin{figure}
\begin{center}
~\epsfig{file=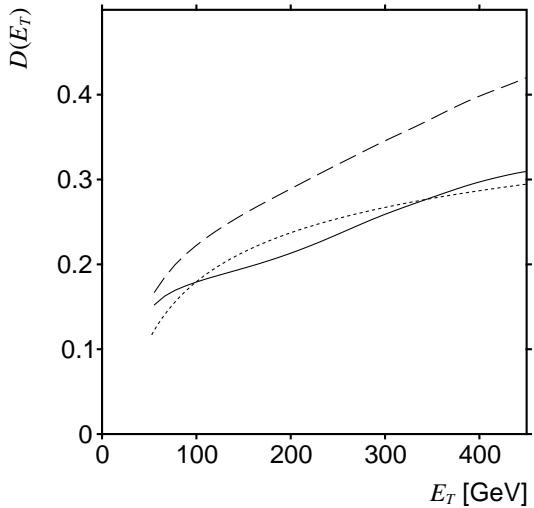,width=3.in,clip=}
\end{center}
\vskip -.7 cm 
\caption[]{
\label{WithGluinoFigure}
The various contributions of gluinos to the single-jet-inclusive cross
section. The dotted line is the effect of only the enhanced
coupling. The dashed line includes the effect of the final state
gluinos and the enhanced coupling.  The solid line includes the
previous gluino effects plus the effect on the initial state parton
distributions; it is therefore our best estimate of the total effect
of the gluino on the cross section.}
\end{figure}

In \fig{WithGluinoFigure} we plot the
%
%
quantity\footnote{ Throughout
  this letter we write ``$d\sigma/dE_T$'' as a shorthand for the
  quantity $\frac{1}{\Delta E_T} \int_{E_T-\Delta E_T/2}^{E_T+\Delta
  E_T/2} (\frac{d\sigma}{dE_T}) \, dE_T$. In the limit $\Delta E_T
  \rightarrow 0$ this is exact. For all of the graphs presented here
  $\Delta E_T = 10$ GeV, a value sufficiently small for our
  purposes. }
$$
D(E_T) = \frac{ d \sigma({\rm with~}\tilde g) / d E_T 
     - d \sigma({\rm without~}\tilde g) / d E_T }
      {d \sigma({\rm without~}\tilde g) / d E_T}\, , 
\equn\label{WithGluinoPlot}
$$
computed with various gluino contributions turned on in the `with
$\gluino$' part.  Since these curves are generated with a leading
order calculation (and because of normalization uncertainties in the
experiments), one must be cautious in interpreting the overall
normalization; however, the shape of the distribution may be expected
to be more robust so that is what we focus on here.

The dotted curve in \fig{WithGluinoFigure} represents the case with a
gluino enhanced coupling constant, but with conventional QCD matrix
elements and parton distributions.  These results have been calculated
using the MRSD0${}^\prime$ \cite{MRSD0} partons, for which the fitted
value of $\alpha_s(M_Z)$ is $0.110\,$.  We use the MRSD0${}^\prime$
partons as a reference point in order to maintain consistency with the
parton distributions of Ref.~\cite{RS} which incorporate gluinos, and
which will be used below.  Observe that although the modified running
of $\alpha_s$ gives an upward tilt to the $E_T$ spectrum, most of the
rise is concentrated in the lower energy range.  We note in passing
that the 17\% enhancement obtained here using two-loop evolution of
$\alpha_s$ is reduced to 14\% if one-loop evolution is used.

The second contribution to the jet cross section comes from the fact
that the gluinos can participate in the hard scattering.  Modern
methods for computing QCD and supersymmetric amplitudes can be found
in Ref.~\cite{ManganoReview}. For completeness, we list the relevant
squared matrix elements (summed over all helicities and colors) as
follows,
$$
\eqalign{
\sum_{\rm spins} & \; \sum_{\rm colors} |{\cal A}_4^{\rm tree} 
                    (1_\qb, 2_q, 3_\gluino, 4_\gluino)|^2 
 = 4 g^4 N_c (N_c^2 -1)  {(t^2 + u^2) \over s^2} \, ,\cr
\sum_{\rm spins} &
                     \;  \sum_{\rm colors} |{\cal A}_4^{\rm tree}
(1_g, 2_g, 3_\gluino, 4_\gluino)|^2\
  = 4 g^4\, {N_c^2  (N_c^2 -1)} \, (u^3 t + t^3 u) \,
\Bigl( {1\over s^2 t^2} + {1\over t^2 u^2} + {1\over s^2 u^2} \Bigr) \;, \cr
 \sum_{\rm spins} & 
                     \;  \sum_{\rm colors} |{\cal A}_4^{\rm tree}
(1_\gluino ,2_\gluino, 3_\gluino, 4_\gluino)|^2
 = 8 g^4 N_c^2 (N_c^2 - 1) \biggl[
  {s^2\over  t^2 u^2} (t^2 - u t + u^2) \cr
& \hskip 6 cm 
+ {t^2  \over s^2 u^2}(s^2 - s u + u^2)
+ {u^2 \over s^2 t^2}( t^2 - s t + s^2)
               \biggr] \, . \cr}
\equn
\label{matrix}
$$
Here,
$$
s = (k_1 + k_2)^2 \, , \hskip 1.5 cm 
t = (k_1 + k_4)^2 \, , \hskip 1.5 cm 
u = (k_1 + k_3)^2 \,  ,
\equn
$$
are the usual Mandelstam variables.  We use the convention that all
particles are in the final state, and the squared matrix elements do
not include phase space symmetry factors.  Since the typical energy
scales probed in jet physics are much higher than the masses of the
individual fermions, it is a reasonable approximation to treat the
lighter quarks and the gluino as massless.  The standard QCD squared
matrix elements may be found, for example, in
Ref.~\cite{PeskinSchroeder}.  (As is usual in jet calculations, we
have ignored the top quark.)

Including the effect of gluinos in the final states together with the
enhanced coupling we obtain the dashed line in \fig{WithGluinoFigure}.
Thus, by ignoring the effect of the light gluino on the initial state
partons we obtain an overall enhancement of the cross section as well
as an overall upward tilt in the $E_T$ spectrum, on the order of 26\%.
Similar upward tilts have been presented in Refs.~\cite{ClavelliET}.

Finally, the gluino also modifies the parton densities of the
proton.  The gluino affects these densities in two ways.  In the DGLAP
equations there are new channels, and there is also an overall
increase in their rate of evolution due to the larger coupling
constant.

The effect of a low mass gluino on parton densities has been
considered in Refs.~\cite{RS,BB,RV}.  A light gluino has a negligibly
small effect on deep inelastic scattering data, including the
kinematic region explored at HERA \cite{RS,RV}.  This is because deep
inelastic scattering is sensitive primarily to the quark
distributions, and these distributions are quite insensitive to the
introduction of a light gluino.  The effect of the gluino on the gluon
distribution is much larger, but deep inelastic scattering is rather
insensitive to the gluon component of the proton.

Since gluons participate directly in the hard scattering at $\bar p p$
colliders (dominating the low $x$ contribution), jet production is
sensitive to the gluon density in the proton.  Modifications to the
gluons, therefore, cannot be ignored.  The solid line in
\fig{WithGluinoFigure} shows the complete effect of including a 5 GeV
gluino; it includes the modified coupling, the new matrix elements and
modified parton distribution functions.  (Below we discuss the effect
of reducing the gluino mass.)  This set of partons\footnote{We thank
  R.G.  Roberts for making these partons available to us.} was
generated by the authors of Ref.~\cite{RS}, by modifying the analysis
used to obtain the standard MRSD0${}^\prime$ structure functions.

The fact that the solid line in \fig{WithGluinoFigure} shows a smaller
rise than the dotted and dashed lines, as we look from low to high
$E_T$, can be traced back to the softening of the gluon and quark
distribution functions in the presence of a gluino.  By softening we
mean that the distribution functions carry less of the proton's
momentum and are shifted to smaller $x$.

The $u$ and $d$ quark densities at moderate $x$ are reduced primarily
as a result of the larger value of $\alpha_s$ in the presence of the
gluino.  This larger value of $\alpha_s$ leads to a more rapid
evolution of the quark densities, which in turn reduces the quark
densities by a few percent at factorization scales $\mu_F\sim 200$ GeV
and $x$ greater than about 0.1.

\begin{figure}
\begin{center}
\vspace{-1cm}
~\epsfig{file=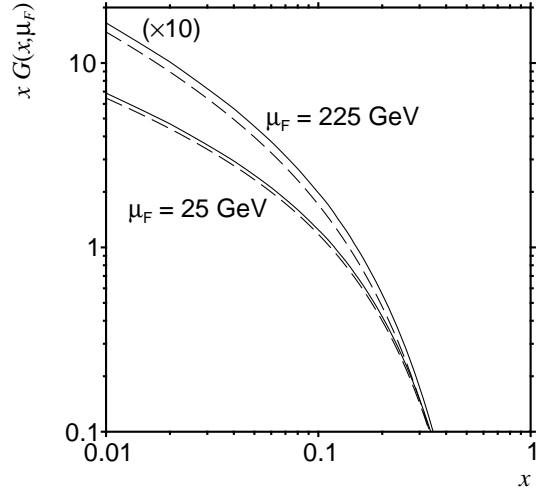,width=3in,clip=}
\end{center}
\vskip -.7 cm 
\caption[]{
\label{SofterPartonsFigure}
The softening of the gluon distribution function due to gluinos.  The
solid lines give the gluon distribution of the MRSD0$^\prime$ partons,
and the dashed lines are the gluon distributions of Ref.~\cite{RS}
which include the effects of a 5 GeV gluino. The distributions are
given for two different factorization scales, $\mu_F$.}
\end{figure}

The gluon distribution is reduced by
the larger value of $\alpha_s$ and the fact that gluons can `split'
into gluinos under the DGLAP evolution.  This leads to a reduction
of about 10\% in the gluon density at factorization scales $\mu_F$ of
order 100 GeV.
In \fig{SofterPartonsFigure} we display the
%
%
gluon momentum distribution with and without a light gluino. We give
the gluon distributions for two factorization scales, $\mu_F = 25$ and
$225$ GeV; these span the scales probed by the Tevatron
single-jet-inclusive experiments.

\begin{figure}
\begin{center}
\vspace{-1cm}
~\epsfig{file=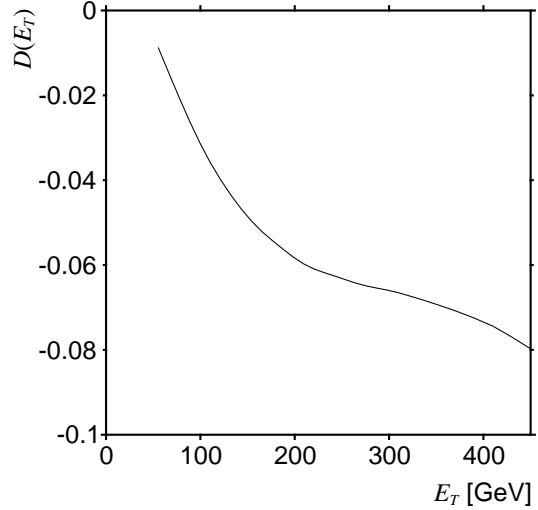,width=3in,clip=}
\end{center}
\vskip -.7 cm 
\caption[]{
\label{NoGluinoFigure}
The fractional decrease in cross section due to the softening of the
gluon distribution of Ref.~\cite{RS} as compared to that of
MRSD0$^\prime$.}
\end{figure}

The effect of this softening on the $E_T$ spectrum is displayed in
\fig{NoGluinoFigure}.  Here we show, as in
%
%
Fig~\ref{WithGluinoFigure}, the fractional change in the jet $E_T$
spectrum using the gluino-modified parton distributions of
Ref.~\cite{RS}, omitting all other effects of the light gluino.  In
other words, we have plotted the standard QCD processes convoluted
with the Standard Model subset of the modified parton distributions,
and adopted the standard (MRSD0$^\prime$) running of the coupling.
This shows the origin of the cancellation seen in
\fig{WithGluinoFigure}: the gluon and quark distributions become
softer.

\begin{figure}
\begin{center}
\vspace{-1cm}
~\epsfig{file=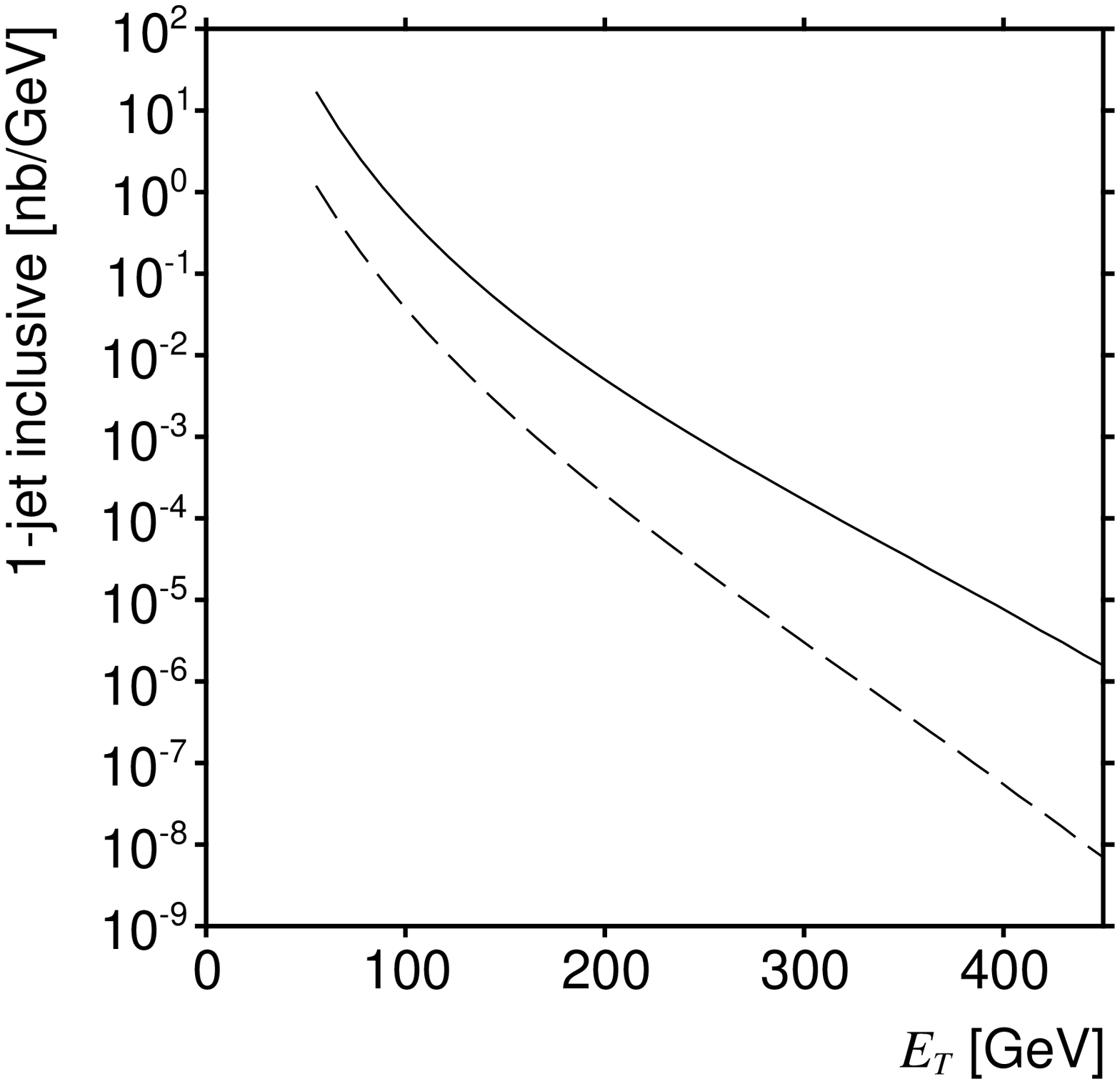,width=3in,clip=}
\end{center}
\vskip -.7 cm 
\caption[]{
\label{GluinoEffectFigure}
The contribution of the initial state gluinos to the cross section is
small, especially in the high $E_T$ region. Here we use the parton
distributions of \cite{RS} breaking the total single-jet rate into two
pieces: the solid curve shows the contribution of the Standard Model
initial states scattering to all final states; the dashed curve is
that for initial states involving one or more gluinos. The quantity
plotted is the differential cross section for the production of
individual jets with a given $E_T$ as a function of $E_T$. It is
further divided by the range in pseudorapidity, $\Delta \eta = 1.2$
(for CDF).}
\end{figure}

Of course, this downward shift in the cross section is reduced
somewhat when initial state gluino scattering is included in the
calculation.  The cross section for gluino scattering is displayed in
\fig{GluinoEffectFigure} for the case of a 5 GeV gluino.  We see that
%
%
this portion of the cross section is significant only at low $E_T$,
and falls off much more rapidly than quark and gluon scattering cross
sections with increasing $E_T$. Thus, in this case initial state
gluinos lead to a negligible increase in the high $E_T$ end of the
spectrum.  The sharp falloff of the gluino contributions may be
understood from the fact that at high $x$ there is little gluino
component to the parton distribution functions; furthermore there is
no $s$-channel scattering of gluinos and quarks, whose distribution is
sizeable at high $x$.

The results presented so far assume a 5 GeV gluino mass, but we are
primarily interested in gluinos with a mass of order 1 GeV.  We expect
that the effects described above will persist for smaller gluino
masses for the following reasons.

Firstly, as we saw above, the introduction of a gluino gives an upward
tilt to the jet $E_T$ spectrum through the modified running of
$\alpha_s$, and the introduction of new final states in quark and
gluon scattering. Both of these effects are largely unchanged
if we reduce the gluino mass from 5 GeV to 1 GeV.

Secondly, we note that the quark distributions are determined by
fitting to deep inelastic scattering data, and are very weakly
modified by the introduction of a light gluino \cite{RS,RV}.  As these
distributions are evolved to higher factorization scales, they are
modified somewhat by the larger value of $\alpha_s$.  However, this
effect is largely independent of the gluino mass.  As a result, a
lower gluino mass can only result in a smaller gluon density inside
the proton, compensated for by a larger gluino density.  In the same
way as was found above, the reduction in the gluon density may be
expected to lessen the upward tilt in the single-jet-inclusive $E_T$
spectrum.  To see this, we note that the parton densities inside the
proton are only logarithmically sensitive to the gluino mass.  For the
enhanced gluino content of the proton to overcome the decrease in the
gluon content, from \fig{GluinoEffectFigure}, we would require as a
lower bound an order of magnitude increase in the high $x$ gluino
distribution.  The relatively weak dependence of the gluino density on
the gluino mass precludes such a large enhancement. As a crude
estimate of the dependence of the gluino density on the gluino mass,
we have evolved the CTEQ3L parton distributions \cite{CTEQ3L} between
factorization scales of 2 and 100 GeV, including gluinos of various
masses.  For gluinos in the 2 to 10 GeV mass range, we see a $\sim \!
40\%$ variation in the gluino density at factorization scales $\mu_F
\sim 100$ GeV.  This mild variation is consistent with the gluino
mass-dependence found in Ref.~\cite{RV}.  Hence, one expects the
cancellation between contributions found above to persist for lower
gluino masses.  Consequently, we put an upper bound of about 16\% on
the expected rise in the cross section between low and high $E_T$ due
to a light gluino.

\section{Relation to Current Experimental Situation}
\label{VsExperiment}

\begin{figure}
\begin{center}
~\epsfig{file=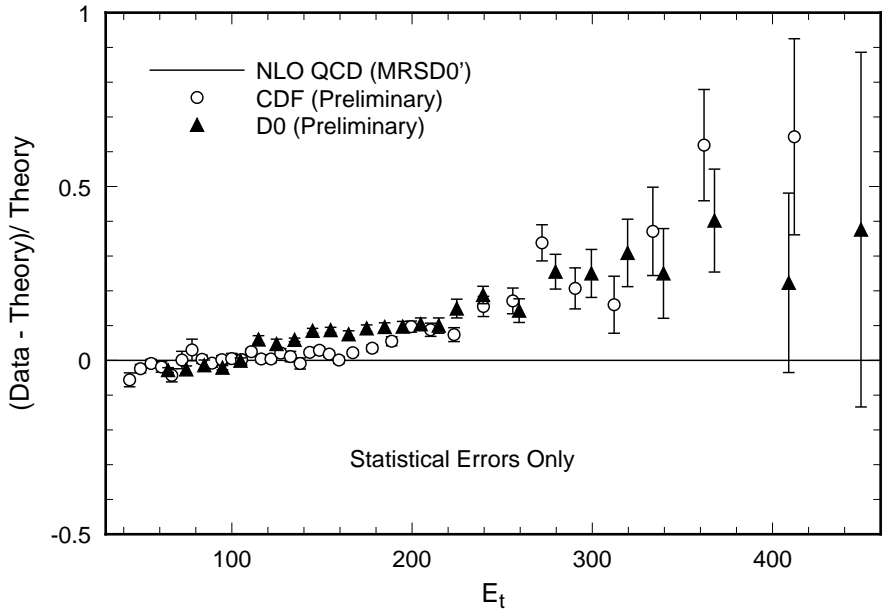,width=5in,clip=}
\end{center}
\vskip -.7 cm 
\caption[]{
\label{CDFdataFigure}
The difference between experiment and NLO theory at 1.8 TeV center-of-mass
using the EKS \cite{EKS} program (from Ref.~\cite{Tung}).}
\end{figure}

The CDF experiment has reported an excess for the single-jet-inclusive
rate at high $E_T$, of the order of 50\%. D0 has not confirmed this
excess, but it has been argued elsewhere \cite{Tung} that, when
analyzed with a common theoretical calculation, the D0 data are not
found to be inconsistent with that of CDF.  In \fig{CDFdataFigure}
%
%
(which is reproduced from Ref.~\cite{Tung}), the observed signal is
seen to be gently rising above the next-to-leading-order Standard
Model theory with increasing $E_T$.

Considering that the cross section falls about seven orders of
magnitude over the $E_T$ range of the plot, the agreement between QCD
and the data is rather impressive.  Nevertheless, the high $E_T$
excess is troublesome. Its interpretation is not clear because of a
variety of experimental and theoretical issues ~\cite{CDFReview}.  On
the theoretical side, for example, it is possible to readjust the
parton distribution functions to remove the excess with only a minor
penalty in $\chi^2$ for the global fit to data
\cite{Tung, CTEQPartons}.

The high $E_T$ rise in the CDF data appears to be of the order of
50\%, which may be contrasted to the smaller $\sim$16\% rise (between
low $E_T$ and high $E_T$) which we found to be due to the light
gluino.  Thus, we conclude that a light gluino is insufficient to
generate the excess observed by CDF.  Note also, from
\fig{WithGluinoFigure}, that much of the increase due to a light
gluino would occur in the lower energy region, which is not apparent
in the CDF data shown in \fig{CDFdataFigure}.

Unfortunately, due to the present uncertainties in both the proton's
gluon distribution \cite{CTEQPartons} and the systematic experimental
uncertainties in the available data, no definitive conclusion is
currently possible as to the origin of the rise.  With smaller
uncertainties it would be possible to rule out or support the
existence of a light gluino, or differentiate between other physics
scenarios.  It would have been desirable to use the data to put bounds
on the appearance of extra light strongly interacting degrees of
freedom; however, remaining uncertainties in the latest parton
distribution functions and the small size of the effect make this task
problematic.

One may, of course, increase the high $E_T$ end of the theoretical
prediction by assuming particular higher energy scale physics
\cite{RandomExplanations}.  Unfortunately, there is little constraint
on how to go about doing this.  Generally, new high scale physics
(with appropriately chosen parameters) increases the cross section
because new channels open.

In principle it should be possible to determine whether the high $E_T$
rise is due to low ($<5$ GeV) or high scale ($>200$ GeV) physics.  Low
energy physics, such as modifications to the parton distributions
\cite{CTEQPartons} or the addition of a light gluino, would require
that the behavior of the cross section at 630 GeV total center of mass
energy be similar to that shown in Fig.~\ref{WithGluinoFigure} for the
1.8 TeV data; essentially one need merely rescale $E_T$ in
Fig.~\ref{WithGluinoFigure} by 630/1800.  The origin of this simple
scaling is that the parton distribution functions undergo little
evolution between the energy scales of interest.  Although the authors
of Ref.~\cite{UA2} have cautioned against direct comparison of
perturbative QCD with their $630$ GeV data, these data have been found
\cite{CTEQPartons} to agree reasonably well with the
next-to-leading-order prediction.  The UA2 data do not show a rise at
high $E_T$.

In addition, we have investigated the effect of a light gluino on the
single-jet-inclusive rate at the $pp$ collider LHC ($\sqrt{s} = 14$
TeV). Over the $E_T$ range 400 GeV to 2 TeV, and with similar cuts on
the pseudorapidity of the jet, we find that the full $D(E_T)$ ratio is
a slowly decreasing function of $E_T$. Over this range in transverse
energy, $D(E_T)$ falls by roughly 7\%.

\section{Summary and Discussion}
\label{Summary}

We have estimated the effect of a light gluino on the
single-jet-inclusive production cross section measured at CDF
\cite{CDF}, D0 \cite{D0}, and UA2 \cite{UA2}.  The gluino affects this
cross section in three ways.  Firstly, the gluino slows the running of
$\alpha_s$, making it larger at high energies.  This acts to enhance
the cross section relative to standard QCD.  Secondly, gluinos can be
pair produced in $\bar q q$ and $g g$ collisions, resulting in a
further enhancement of the cross section. (These two contributions
have very recently been discussed in Ref.~\cite{ClavelliET}.)
Finally, the gluino modifies the DGLAP evolution of the parton
densities of the proton through the introduction of new states as well
as the value of $\alpha_s$.  We have found that this last effect
cancels against the first two, leaving no more than a $16\%$
enhancement between the low and high $E_T$ ends of the spectrum for
the case of a 5 GeV gluino.  We have argued that this cancellation
should persist for smaller gluino masses.

The current experimental uncertainties, and the
uncertainties in the parton densities \cite{CTEQPartons}, make it 
unlikely that single-jet-inclusive data will be useful to
differentiate between the Standard Model and supersymmetric models
with a light gluino.

\section*{Acknowledgments}

We thank R.D. Peccei for encouraging us to write this paper.  We thank
E.W.N. Glover for providing us with a a copy of JETRAD as well as very
helpful comments.  We also thank both W.T. Giele and D.A. Kosower for
helpful conversations.  We thank W.J. Stirling, R.G. Roberts and
W.K. Tung for correspondence concerning parton distributions.  This
work was supported in part by the Department of Energy under Grant
FG03-91ER40662 and by the Alfred P. Sloan Foundation under grant
BR-3222.


\begin{thebibliography}{99}

\bibitem{mayberefs}
S. Dawson, E. Eichten and C. Quigg, Phys.\ Rev.\ {\bf D31} (1985) 1581;\\
L. Clavelli, Phys. Rev. {\bf D46} (1992) 2112;\\
J. Ellis, D. Nanopoulos, and D. Ross, Phys. Lett. {\bf B305} (1993) 375;\\
L. Clavelli, P. Coulter and K. Yuan, Phys. Rev. {\bf D47} (1993) 121;\\
R. Mu\~noz-Tapia and W.J. Stirling, Phys. Rev. {\bf D49} (1994) 3763;

\bibitem{Farrar}
G.R. Farrar Phys.\ Rev.\ {\bf D51} (1995) 3904;
preprint hep-ph/9504295;
preprint hep-ph/9508291;
preprint hep-ph/9508292.

\bibitem{bmdmp}
R.C. Ball {\it et al.}, Phys. Rev. Lett. {\bf 53} (1984) 1314;\\
Charm Collaboration, F. Bergsma {\it et al.}, Phys. Lett. {\bf 121B}
(1983) 429;\\
WA66 Collaboration, A. M. Cooper-Sarkar {\it et al.},
Phys. Lett. {\bf 160B} (1985) 212.

\bibitem{stable}
R. Arnold {\em et al.}, Phys. Lett. {\bf 186B} (1987) 435;\\
NA3 Collaboration, J. Badier {\it et al.}, Zeit. Phys. {\bf C31}, 21
(1986);\\
R.H. Bernstein {\it et al.}, Phys. Rev. {\bf 37D} (1988) 3103;\\
M. Borquin {\em et al.}, Nucl. Phys. {\bf B153} (1979) 13;\\
D. Cutts {\it et al.}, Phys. Rev. Lett. {\bf 41} (1978) 363;\\
H.R. Gustafson {\it et al.}, Phys. Rev. Lett. {\bf 37} (1976) 474;\\
T.T. Nakamura {\it et al.}, Phys. Rev. {\bf D39} (1989) 1261.

\bibitem{Upsilon}
ARGUS Collaboration, H. Albrecht {\it et al.}, Phys. Lett. {\bf 167B} 
(1986) 360; \\
CUSB Collaboration, P. M. Tuts {\it et al.}, Phys. Lett. {\bf 186B}
(1987) 233.

\bibitem{UA1}
UA1 Collaboration, C. Aljabar {\it et al.}, Phys. Lett. 198B, 261 (1987).

\bibitem{Fupsilon}
G. R. Farrar, preprint hep-ph/9603271.

\bibitem{E761}
I.F.Albuquerque {\it et al.}, (E761 Collaboration), preprint hep-ex/9604002, 
submitted to Phys.\ Rev.\ Lett.

\bibitem{RS}
R.G. Roberts and W.J. Stirling, Phys.\ Lett.\ {\bf B313} (1993) 453.

\bibitem{RV}
R. R\"uckl and A. Vogt  Z. Phys. {\bf C64} (1994)
431  (hep-ph/9404355).

\bibitem{BB}
J. Bl\"umlein and J. Botts, Phys. Lett. {\bf 325B} (1994) {190}; 
Phys. Lett. {\bf 331B} {450} (1994).

\bibitem{ClavelliDiJet}
I. Terekhov and L. Clavelli, preprint hep-ph/9603390.

\bibitem{CDF2JET} CDF Collaboration, R.M. Harris {\em for the
 collaboration} Fermilab-CONF-95/152-E.

\bibitem{GieleGlover}
W.T. Giele and E.W.N. Glover, Phys.\ Rev.\ {\bf D46} (1992) 1980. 
W.T. Giele, E.W.N. Glover and D.A. Kosower, 
Nucl.\ Phys.\ {\bf B403} (1993) 633;
Phys.\ Rev.\ Lett.\ {\bf 73} (1994) 2019. 

\bibitem{GMRS}
E.W.N. Glover, A.D. Martin, R.G. Roberts, W.J. Stirling, 
preprint hep-ph/9603327.

\bibitem{DGLAP}
Yu. L. Dokshitzer, Sov. Phys. JETP {\bf 46} (1977) 641;\\
V.N. Gribov and L.N. Lipatov, Sov. J. Nucl. Phys. {\bf 15}
(1972) 675; \\
G. Altarelli and G. Parisi, Nucl.\ Phys.\ {\bf B126} (1977) 298;\\
G.\ Curci, W.\ Furmanski and R.\ Petronzio, Nucl.\ Phys.\ {\bf B175}
(1980) 27.

\bibitem{DIS}
See, e.g., I. Hincliffe, in The Review of Particle Properties,
M. Aguilar-Benitez {\it et al.}, Phys. Rev. {\bf D50} (1994) 1297.

\bibitem{MRSD0}
 A.D. Martin, W.J. Stirling and R.G. Roberts,
Phys. Lett. {\bf B306} (1993) 145; {\em Erratum ibid,} {\bf B309}
(1993) 492.

\bibitem{ManganoReview}
M. Mangano and S.J. Parke, Phys.\ Rep.\ {\bf 200} (1991) 301;\\
L. Dixon, preprint hep-ph/9601359, to appear 
in {\it Proceedings of Theoretical Advanced Study Institute in 
Elementary Particle Physics (TASI 95)}, ed.\ D.E.\ Soper.

\bibitem{PeskinSchroeder}
M.E.\ Peskin and D.V.\ Schroeder, {\it An Introduction to Quantum Field Theory}
(Addison-Wesley, 1995).

\bibitem{ClavelliET}
A.K. Grant, talk presented at
``PHENO '96: Recent Developments in Phenomenology'', University of
Wisconsin, Madison, April 1-3, 1996;\\
L. Clavelli and I. Terekhov, preprint hep-ph/9605463.

\bibitem{CTEQ3L} 
H.L. Lai {\it et al.}, Phys. Rev. Lett. {\bf 75} (1995) 608.

\bibitem{Tung}
H.L. Lai and W.K. Tung, preprint hep-ph/9605269.

\bibitem{EKS}
S.D.\ Ellis, Z. Kunszt and D.E.\ Soper,
Phys.\ Rev.\ {\bf D40} (1989) 2188 ;
Phys.\ Rev.\ Lett.\ {\bf 64} (1990) 2121;
Phys.\ Rev.\ Lett.\ {\bf 69} (1992) 1496.

\bibitem{CDFReview}
CDF Collaboration, A. Bhatti {\em et al.},
preprint  FERMILAB-CONF-95/192-E.

\bibitem{CTEQPartons}
J. Huston, {\it et al.}, preprint hep-ph/9511386.

\bibitem{RandomExplanations}
R.S. Chivukula, A.G. Cohen and E.H. Simmons, preprint hep-ph/9603311;\\
B.A. Arbuzov, preprint hep-ph/9602416; \\ 
M. Bander, preprint hep-ph/9602330.

\bibitem{UA2}
UA2 Collaboration, J. Alitti {\it et al.}, Phys. Lett. {\bf 257B} (1991) 232.

\bibitem{CDF}
CDF Collaboration, F. Abe {\it et al.},  preprint hep-ex/9601008.

\bibitem{D0}
D0 Collaboration, G. Blazey {\it et al.}, Proceedings of
the XXXI Rencontres de Moriond (March 1996).

\end{thebibliography}
\end{document}